# Space-division multiplexing for fiber-wireless communications

**Ivana Gasulla, *Member, IEEE*, Sergi García, David Barrera, Javier Hervás and Salvador Sales, *Member, IEEE***
*ITEAM Research Institute, Universitat Politècnica de València, 46022 Valencia, Spain*
*e-mail: ivgames@iteam.upv.es*

**ABSTRACT**
We envision the application of optical Space-division Multiplexing (SDM) to the next generation fiber-wireless communications as a firm candidate to increase the end user capacity and provide adaptive radiofrequency-photonic interfaces. This approach relies on the concept of "fiber-distributed signal processing", where the SDM fiber provides not only radio access distribution but also broadband microwave photonics signal processing. In particular, we present two different SDM fiber technologies: dispersion-engineered heterogeneous multicore fiber links and multicavity devices built upon the selective inscription of gratings in homogenous multicore fibers.

## 1. INTRODUCTION

The upcoming 5G communications scenario will require evolving existing wireless radio access platforms in combination with new photonics technologies to address the current limitations to massive capacity, connectivity and flexibility [1]. Broad bandwidth, low latency and extreme reliability will be essential and require a full convergence between the optical fiber and the wireless network segments. Convergence to be sought in terms not only of radio access *signal distribution* between a central office and different remote locations, (including Multiple Input Multiple Output (MIMO) antenna connectivity), but also of radiofrequency (RF) *signal processing* [2,3]. We have proposed the use of multicore (MCF) optical fibers [4] as a compact media to provide both distribution and processing functionalities "simultaneously", leading to the concept of "fiber-distributed signal processing" [5-7]. Many of the signal processing functionalities, such as microwave filtering, radio beam-steering and signal generation, rely on a key building block: the true time delay line (TTDL), [3]. This subsystem provides a frequency independent and tunable group delay within a given frequency range. Fig. 1 shows the general idea underlying the TTDL built upon a generic MCF. There, we obtain at the output of the MCF medium a set of different time-delayed samples of the RF signal characterized by a constant basic differential delay $\Delta\tau$ between adjacent samples. This scheme corresponds to 1D (1-dimensional) operation since the $N$ samples are achieved by exploiting the spatial diversity behavior of the MCF. However, this TTDL potentially offers not only 1D, but also 2D operation if we incorporate the optical wavelength diversity provided, for instance, by an array of lasers emitting at different wavelengths. Actually, MCFs can be classified as homogeneous or heterogeneous. In the case of a heterogeneous MCF, we can design each core with a different refractive index profile as to provide a different group delay, [5-7]. In the case of homogeneous MCFs, where all the cores are in principle identical, we must incorporate dispersive optical elements in each one of the cores, [7].

We show in this paper how we can implement a sampled TTDL in both dispersion-engineered heterogeneous MCFs and homogeneous MCFs with inscribed dispersive elements. As a proof of concept, we will investigate the application of the developed delay lines to one of the signal processing functionalities that will be demanded in converged fiber-wireless communications: reconfigurable microwave signal filtering.

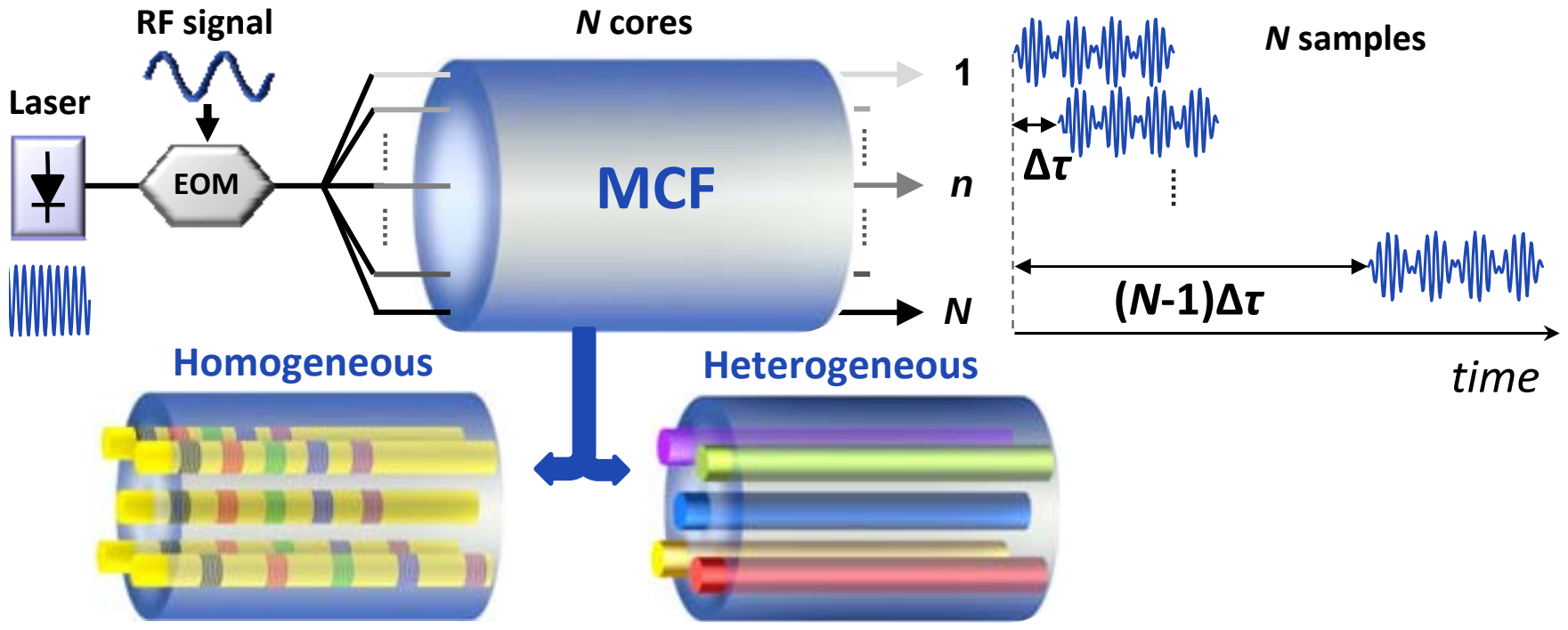

*Figure 1. General concept underlying the sampled true time delay line built upon a MCF medium*

## 2. DISPERSION-ENGINEERED HETEROGENEOUS MULTICORE FIBERS

The design of heterogeneous MCFs to operate as group-index-variable delay lines gives us the possibility of implementing fiber-distributed signal processing elements with a length up to a few kilometers where we can process the data signal while transmitting it. A group-index-variable TTDL, as opposed to length-variable TTDLs, involves a variation in the propagation velocity of the optical fiber cores or waveguides involved. The design of a heterogeneous MCF to behave as a group-index-variable delay line implies that each core features an independent group delay with a linear dependence on the optical wavelength. Actually, we can expand the group delay $\tau_{g,n}(\lambda)$ of a particular core $n$ as a 3rd-order Taylor series around an anchor wavelength $\lambda_0$ as

$$\tau_{g,n}(\lambda) = \tau_g(\lambda_0) + D_n \cdot (\lambda - \lambda_0) + \frac{S_n}{2} \cdot (\lambda - \lambda_0)^2, \tag{1}$$

where $D_n$ is the chromatic dispersion and $S_n$ the dispersion slope of core $n$. For proper TTDL operability, $D_n$ must increase linearly with the core number while we assure a linear behavior of the group delay with the optical wavelength within the optical operation range. In addition, we must assure lower levels of intercore crosstalk and robustness against fiber bend curvatures [10].

Although this delay line offers 2D operation by using either the spatial or the wavelength diversities, we will focus for simplicity on the spatial diversity, where the basic differential delay $\Delta\tau_{n,n+1}$ between samples is given by the propagation difference created between two adjacent cores $n$ and $n$+1 for a particular wavelength $\lambda_m$:

$$\Delta\tau_{n,n+1}(\lambda_m) = \Delta D \cdot (\lambda_m - \lambda_0) + \frac{\Delta S}{2} \cdot (\lambda_m - \lambda_0)^2, \tag{2}$$

being $\Delta D = D_{n+1} - D_n$ the common incremental dispersion between adjacent cores that must be kept constant, and $\Delta S = S_{n+1} - S_n$ the dispersion slope variation that we must minimize to reduce the quadratic wavelength dependence.

We have designed and evaluated several heterogeneous MCF links attending to different delay line and transmission performance requirements, [5,6]. For instance, the table in Fig. 2(a) gathers the design parameters for the 7 trench-assisted refractive index profiles of a 7-core fiber (35-μm core pitch and 125-μm cladding diameter) that is optimized in terms of both higher-order dispersion and intercore crosstalk: core radius $a_1$, core-to-cladding relative index difference $\Delta_1$, core-to-trench separation $a_2$, trench width $w$ and effective index $n_{eff}$. The set of dispersion parameters $D_n$ range from 14.75 up to 20.75 ps/(km·nm) with an incremental dispersion $\Delta D$ = 1 ps/(km·nm). The fiber satisfies as well the common group index condition at $\lambda_0$ = 1550 nm. The cladding-to-trench relative index is $\Delta_2 \approx$ 1% in all cores. Figure 2(b) shows the computed group delay for each core as a function of the optical wavelength where we observe, as requested for TTDL operation, that all the cores share a common group delay at $\lambda_0$ and we obtain linearly incremental group delay slopes. Actually, for a fixed wavelength (for instance $\lambda$ = 1560 nm), the resulting TTDL is characterized by a basic differential delay $\Delta\tau_{n,n+1}$ lower than the one obtained for a higher wavelength (for instance $\lambda$ = 1570 nm).

As a proof of concept, we evaluated the performance of the designed TTDL when it is applied to reconfigurable microwave signal filtering, [3]. The filtering effect is achieved by combining and photodetecting all the delayed RF signal samples coming from the MCF output, as illustrated in Fig. 1. The Free Spectral Range (FSR) of the filter is then given by FSR = $1/\Delta\tau$ and can be controlled by tuning the optical wavelength of the laser if we operate the TTDL using the spatial diversity provided by the cores. Figure 2 (c) shows the computed transfer function of the microwave filter as a function of the RF frequency for a 10-km MCF link and two different optical wavelengths. We can see how the increase of the optical wavelength from 1560 to 1570 nm reduces the FSR from 10 to 5 GHz without any filter degradation due to higher-order dispersion effects. Other Microwave Photonics functionalities, as optical beamforming for phased array antennas, were also evaluated in [6].

We cannot forget that one of the major detrimental effects in heterogeneous MCF transmission arises from the crosstalk dependence on the phase-matching condition between adjacent cores when the fiber is bent [10]. To overcome this, we must optimize our designs by maximizing, as much as possible, the effective index difference $\Delta n_{eff}$ between adjacent cores as to improve the threshold bending radius $R_{pk}$. The MCF described by the parameters gathered in Fig. 3(a) presents an $R_{pk}$ close to 103 mm and a worst-case crosstalk above the phase-matching region below -80 dB, [6].

a) *Core properties*

| Core $n$ | $a_1$ (μm) | $\Delta_1$ (%) | $a_2$ (μm) | $w$ (μm) | $n_{eff}$ |
|---|---|---|---|---|---|
| 1 | 3.42 | 0.3864 | 5.48 | 3.02 | 1.453384 |
| 2 | 3.60 | 0.3762 | 5.03 | 2.61 | 1.453465 |
| 3 | 3.62 | 0.3690 | 4.35 | 3.32 | 1.453386 |
| 4 | 4.26 | 0.3588 | 4.92 | 4.67 | 1.453881 |
| 5 | 3.49 | 0.3476 | 2.81 | 5.41 | 1.452878 |
| 6 | 4.79 | 0.3435 | 3.35 | 3.32 | 1.454041 |
| 7 | 4.98 | 0.3333 | 2.42 | 4.05 | 1.453979 |

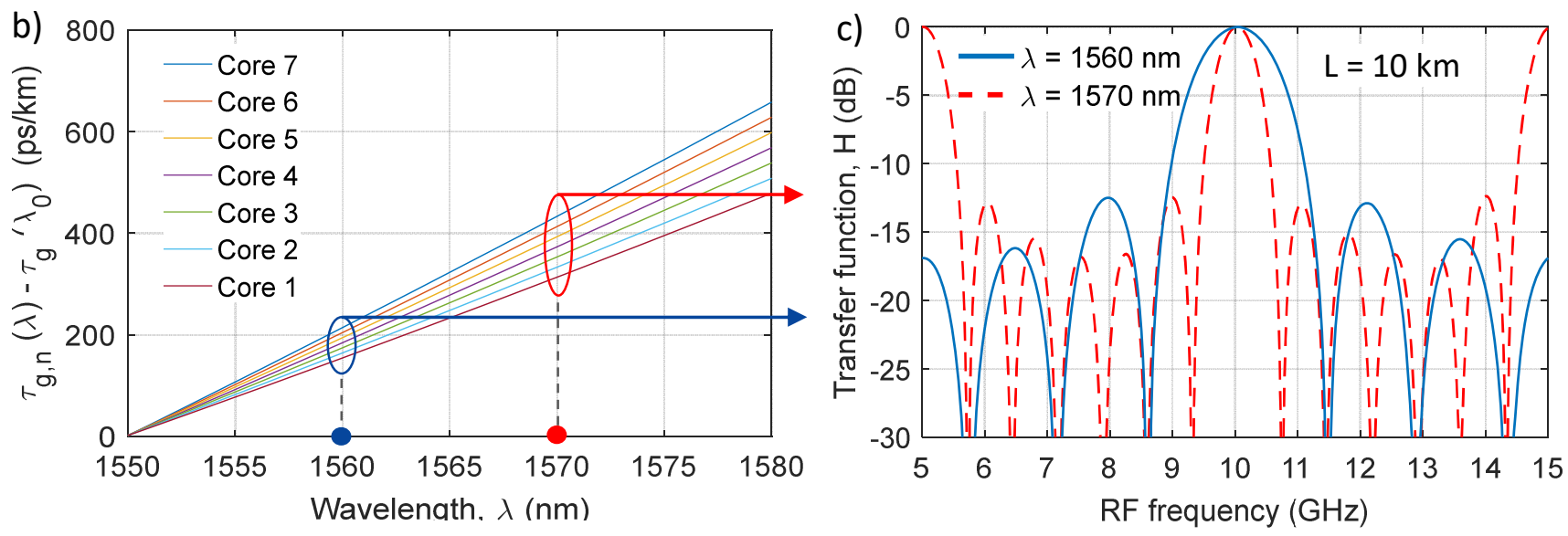

*Figure 2. (a) Core design parameters for the heterogeneous MCF; (b) Computed group delays versus wavelength and (c) Computed transfer function of the microwave signal filter when operating in spatial diversity for L=10 km and wavelengths 1560 and 1570 nm.*

## 3. HOMOGENOUS MULTICORE FIBERS WITH SELECTIVE FIBER BRAGG GRATINGS

An alternative solution is to use commercial homogeneous MCFs, where all the cores are -in principle- identical and share the same group delay. This involves the design of a multicavity device where the diversity in the propagation characteristics comes from the introduction of series of fiber Bragg gratings (FBGs) at *selective* positions along the MCF cores. The inscription of FBGs in singlemode fibers has been widely investigated as a dispersive 1D sampled delay line [8], where the use of the optical wavelength diversity allows the tunability of the basic differential delay. If we incorporate the spatial diversity provided by the cores, we obtain a 2D TTDL that offers a higher degree of flexibility, versatility and compactness.

Previous work on FBG inscription in MCFs has focused mainly on the *simultaneous* writing of the same grating (i.e. placed at the same position) in all the cores for sensing and astrophotonics [9]. Seeking for the mentioned *selective* inscription, we have developed the first fabrication method to inscribe high-quality gratings characterized by arbitrary frequency spectra and located in arbitrary longitudinal positions along the individual cores of a MCF, [7]. Based on the moving phase mask technique, the inscription setup allows to inscribe in different groups of cores, including individual cores, by controlling the shape and position of the laser beam, [7]. The beam is conditioned by a mirror mounted in a piezo-electric transducer that can induce a beam vertical deflection and by two cylindrical lenses that control the beam height and width. Actually, the dimension and cross-sectional arrangement of the cores determine the height, which was confined to a range from 30 to 50 µm. Furthermore, we adjust the beam width and the distance of the fiber to the phase mask to control the interference pattern created after the phase mask. For a phase mask with a period of 1070 nm and a fiber placed as close as possible to the phase mask, the maximum width to inscribe in an individual outer core is approximately 23 µm.

One of the fabricated multicavity devices is based on the selective inscription of gratings in three outer cores. In each of these cores, we inscribed an array of three equally-spaced uniform gratings, being each one located in a different longitudinal position and centered at a different Bragg wavelength. The left part of Fig. 3(a) shows the cross-sectional view of the MCF (7-core fiber with a cladding diameter of 125 µm and a pitch of 35 µm, from Fibercore) where the three target cores are encircled. The right part of Fig. 3(a) shows the scheme of the FBGs inscribed in cores 4, 5, and 6, where we bring attention to the delay line duality obtained by exploiting both the spatial and the wavelength diversities. To assure a proper wavelength diversity operation, the 3 arrays of FBGs have been written to feature incremental distances between FBGs from core to core, i.e., 20-mm, 21-mm and 22-mm distances, respectively, within cores 6, 5 and 4. On the other hand, the spatial diversity arises from the incremental displacement applied between FBGs centered at the same wavelength but inscribed in adjacent cores, i.e., 6-mm, 7-mm and 8-mm displacements, respectively, for the wavelengths $\lambda_1$ = 1537.07 nm, $\lambda_2$ = 1541.51 nm and $\lambda_3$ = 1546.26 nm. Figure 3(b) shows the optical spectrum in reflection for cores 4, 5 and 6. We see small fluctuations in the strength of the inscribed gratings with a maximum difference of 3 dB. In the other outer cores (cores 2, 3 and 5), only residual gratings were inscribed with an optical power peak 15 to 29 dB lower than those of cores 4, 5 and 6. We must note that the central core received radiation from the three inscription processes, resulting in a maximum peak interference level 8 dB lower.

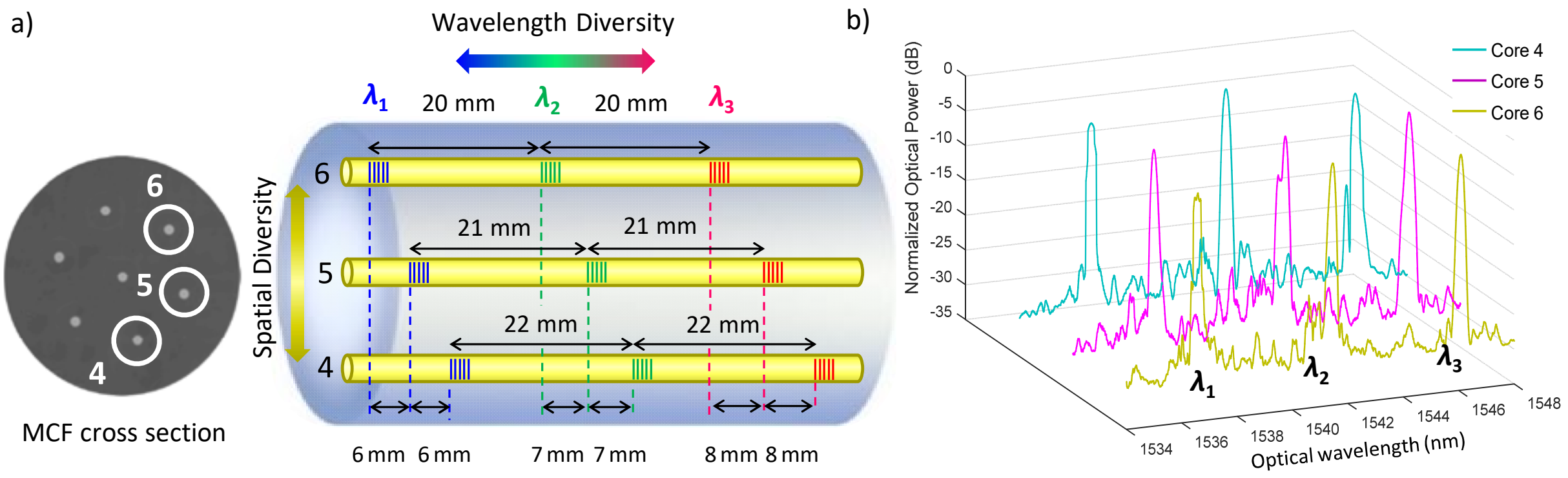

*Figure 3. (a) Cross section of the homogeneous MCF and schematic of the FBG selectively inscribed in cores 4, 5 and 6; (b) Measured optical spectra in reflection for the target cores.*

We experimentally evaluated the performance of the fabricated device in the context of reconfigurable microwave signal filtering. Figure 4(a) depicts the experimental setup where we feed the TTDL by either an array of three low-linewidth lasers or a broadband optical source followed by a 2-nm-bandwidth filter. The broadband source is required to avoid coherent interference when we operate in spatial diversity as we detect together the samples originated from different cores with $\Delta\tau$ being lower than the coherence time of the lasers. Figure 4(b) shows the measured electrical frequency responses for the 3-tap filters obtained when we operate in wavelength diversity, i.e. gathering the delayed samples coming from a given core. We see how we can reconfigure the filter, varying the FSR from 4.45 up to 4.97 GHz. The same device can operate using the spatial diversity provided by the cores, i.e. gathering the delayed samples from different cores that correspond to the same optical wavelength. Figure 4(c) shows the 3-tap filter responses given by the three input optical wavelengths, where we can see how the FSR varies from 12.50 up to 17.76 GHz. These experimental results are in good agreement with the ones expected from the theory, with small discrepancies in the basic differential delays that may be caused by discrepancies between the theoretical and actual values of the core refractive indices and small imbalances in the reflectivity strength of the FBGs.

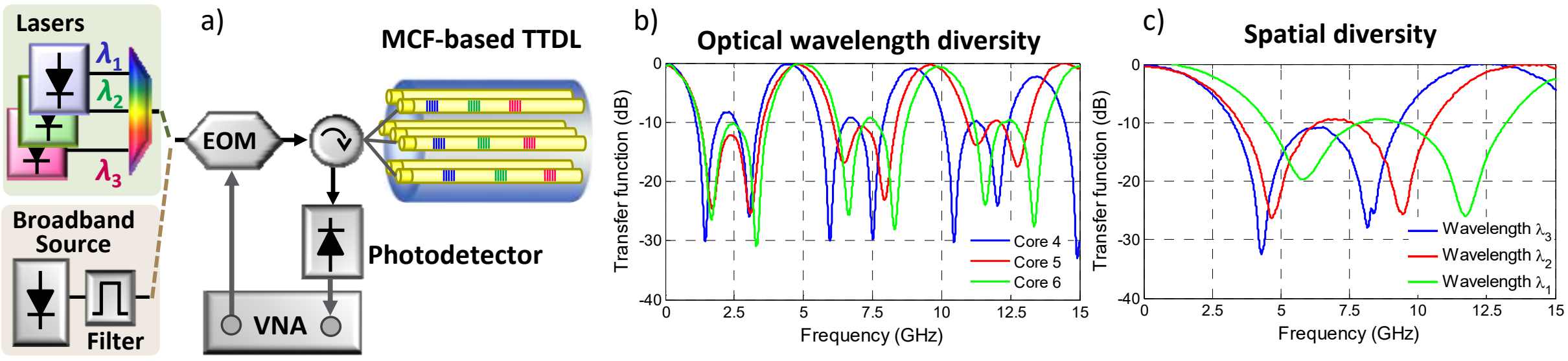

*Figure 4. (a) Microwave filter experimental setup; measured 3-tap filter transfer functions using (b) wavelength diversity (for each one of the target cores) and (c) spatial diversities (for each one of the optical wavelengths).*

## 4. CONCLUSIONS

Space-division Multiplexing technologies in optical fibers can be exploited beyond data signal distribution in high-capacity digital communications systems. We demonstrated that they can serve as a compact fiber medium to provide "fiber-distributed signal processing" in the next generation of fiber-wireless communications, as 5G radio access networks and the Internet of Things. In particular, multicore fibers can be designed to act as sampled true time delay lines for radiofrequency signals. In the case of heterogeneous multicore fibers, we showed that the proper design of the trench-assisted refractive index profile of each core results in length-distributed true time delay lines (up to a few km) with a linear tunability with the operational optical wavelength. In the case of homogeneous multicore fibers, the incorporation of FBGs placed at different positions along the individual cores of the fibers allows the implementation of compact MCF-based devices with a length lower than 10 cm. These fiber delay line elements open the way towards the development of fiber-distributed signal processing for microwave and millimeter-wave signals in a single fiber where a variety of functionalities can be implemented, including arbitrary waveform generation, microwave signal filtering, optical beamforming networks for phased array antennas, MIMO antenna connectivity and analogue-to-digital conversion.

## ACKNOWLEDGEMENTS

This research was supported by the ERC Consolidator Grant 724663, the Spanish Projects TEC2014-60378-C2-1-R, TEC2015-62520-ERC and TEC2016-80150-R, the Spanish MECD FPU scholarship (FPU13/04675) for J. Hervás, the Spanish MINECO (BES-2015-073359) scholarship for S. García, and the Spanish MINECO Ramón y Cajal Program (RYC-2014-16247) fellowship for I. Gasulla.

## REFERENCES

[1] Samsung Electronics Co, "5G Vision", available at http://www.samsung.com/global/business-images/insights/2015/Samsung-5G-Vision-0.pdf, 2015.

[2] J. Yao: Microwave photonics, *J. Lightwave Technol.*, vol. 27, pp. 314-335, 2009.

[3] J. *Capmany et al.*: Microwave photonic signal processing, *J. Lightwave Technol.,* vol. 31, pp. 571-586, 2013.

[4] D. J. Richardson, J. M. Fini and L. E. Nelson: Space-division multiplexing in optical fibres, *Nat. Photonics*, vol. 7, pp. 354-362, 2013.

[5] S. Garcia and I. Gasulla: Design of heterogeneous multicore fibers as samples true-time delay lines, *Opt. Lett.*, vol. 40, pp. 621-624, 2015.

[6] S. Garcia and I. Gasulla: Dispersion-engineered multicore fibers for distributed radiofrequency signal processing, *Opt. Express*, vol. 24, pp. 20641-20654, 2016.

[7] I. Gasulla, D. Barrera, J. Hervás and S. Sales: Spatial Division Multiplexed Microwave Signal processing by selective grating inscription in homogeneous multicore fibers, *Sci. Reports*, no. 41727, pp.1 -10, 2017.

[8] C. Wang and J. Yao: Fiber Bragg gratings for microwave photonics subsystems, *Opt. Express*, vol. 21, pp. 22868-22884, 2013.

[9] T. Birks *et al.*: Multicore optical fibres for astrophotonics, in *CLEO/Europe and EQEC 2011 Conference Digest*, Munich, Germany, 2011, paper JSIII2_1.

[10] J. Tu *et al.*: Optimized Design Method for Bend-Insensitive Heterogeneous Trench-Assisted MultiCore Fiber with Ultra-Low Crosstalk and High Core Density, *J. Lightwave Technol.*, vol. 31, pp. 2590-2598, 2013.